\documentclass{agnspec}
\newcommand{\etal}{et al.}
\newcommand{\fe}{Fe~K$\alpha$}
\newcommand{\kev}{keV}
\newcommand{\mcg}{MCG--6-30-15}
\usepackage{graphics}
\usepackage{psfig}

\begin{document}
\title{Reflection Spectra from Photoionized Accretion Discs}
\author{D.R. Ballantyne\inst{1}, R.R. Ross\inst{2} \and A.C. Fabian\inst{1}}
\institute{Institute of Astronomy, University of Cambridge, Cambridge,
U.K. CB3 0HA
\and  Physics Department, College of the Holy Cross, Worcester, MA
01610, USA}
\maketitle

\begin{abstract}
We review recent progress on the modeling and use of reflection
spectra from irradiated and ionized accretion discs. On the
computational side, calculations of reflection spectra from discs
with non-uniform density structure have shown that thermal
instabilities can effect the predictions. Ionized reflection spectra
have been used effectively in fitting data of Narrow-Line Seyfert 1
galaxies, and have placed constraints on the strength and shape of soft
X-ray emission lines.
\end{abstract}

\section{Introduction}
\label{sect:intro}
With the discovery of \fe\ emission at 6.4~\kev\ and spectral
hardening at 20--30~\kev\ in the X-ray spectra of Seyfert~1 galaxies
(Pounds \etal\ 1990), it was realized that there is a significant
amount of reprocessing material in the vicinity of the X-ray
source. This material must be optically thick to electron scattering
and relatively 'cold' (as compared to the X-ray emitting plasma). The
observation of the relativistically broadened \fe\ line in \mcg\ by
Tanaka \etal\ (1995) confirmed that, at least in some Seyfert~1
galaxies, the reflecting material was the inner part of the accretion
disc. Therefore, the possibility arises of unraveling some of the
mysteries of AGN accretion flows, such as its geometry, by studying
the reflection signatures in X-ray spectra.

\section{Constant density models}
\label{sect:cdens}
Numerical models of reflection spectra from accretion discs began with
the simplest case: assuming a static, neutral and constant density
slab of material irradiated by a power-law continuum of X-rays. George
\& Fabian (1991) and Matt, Perola \& Piro (1991) performed Monte-Carlo
calculations of the reflection spectra in such circumstances, and
provided predictions on the equivalent width (EW) of \fe\ for
different reflection geometries.

\begin{figure}
\centerline{\psfig{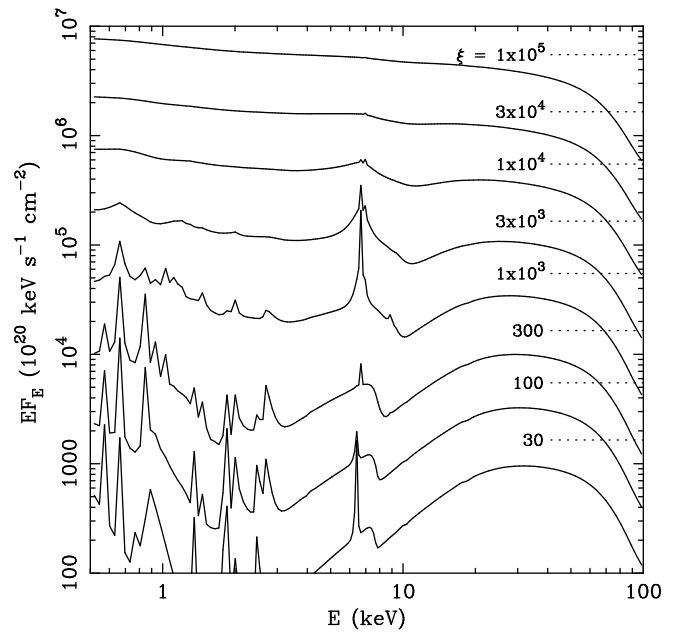}}
\caption[]{Reflection spectra (solid lines) from constant density
slabs illuminated by a power-law continuum of X-rays with photon-index
$\Gamma=2$ (dashed lines). The ionization parameter $\xi$ of the
reflector increases from 30 to $10^5$ going from bottom to top in the
plot. Taken from Ross, Fabian \& Young (1999).}
\label{fig:rfy}
\end{figure}

Of course, assuming that the gas remains neutral while it is being
bombarded by X-rays is not very satisfying, and this led Ross \& Fabian
(1993) to model photoionized reflectors (see also
\.{Z}ycki \etal\ 1994). As might easily be imagined, allowing the gas
to be ionized changes significantly the types of features imprinted on
the X-ray reflection spectrum. This is illustrated in
Figure~\ref{fig:rfy} taken from the paper by Ross, Fabian \& Young
(1999) where extensive calculations are presented. Here we see X-ray
reflection spectra for different cases of the ionization parameter
$\xi = 4 \pi F_{\mathrm{X}} / n_{\mathrm{H}}$ where $F_{\mathrm{X}}$
is the total flux incident on the slab and $n_{\mathrm{H}}$ is the
hydrogen number density of the gas.
In every case the reflector is irradiated by a power-law continuum
with photon-index $\Gamma=2$ (denoted by the dotted lines in
Fig.~\ref{fig:rfy}). When the ionization parameter is small the gas is
weakly ionized and there is a large amount of photoelectric
absorption by metals between 0.1 and 20~\kev. This absorbed energy is
thermalized by the gas and re-radiated in the soft X-rays and
extreme-UV wavebands (not shown in Fig.~\ref{fig:rfy}). The same
transitions which can absorb X-ray photons can also emit them, so the
recombination and fluorescent lines of the abundant metals also occur
between 0.1 and 20~\kev. Due to its relatively high cosmic abundance
and large fluorescent yield the \fe\ line at 6.4~\kev\ is the most
prominent of these lines. Above $\sim$20~\kev\ electron
scattering results in the albedo of the reflector to approach unity, until
above $\sim$50~\kev\ where electron recoil causes it to decrease
again. 

Increasing the ionization parameter of the slab removes the ability of
the gas to absorb the X-ray photons, and so increases the albedo of
the reflector. This occurs first at lower energies and then moves to
higher energies as the gas becomes more ionized. Therefore, the \fe\
line is a strong feature in the reflection spectrum over more than three orders
of magnitude in ionization parameter. The line does shift energy,
however, from 6.4~\kev\ when Fe is only weakly ionized to 6.7~\kev\
when Fe is helium-like, having a significant impact on its observed
strength (Matt, Fabian \& Ross 1993, 1996). One other important effect
of a larger ionization parameter is the broadening of the spectral
features by Compton scattering (this can be especially seen in the
$\xi=3\times 10^3$ model in Fig.~\ref{fig:rfy}).

Constant density reflection models such as the ones shown in
Figure~\ref{fig:rfy} are useful because their properties are easily
parameterized by only two parameters ($\xi$ \& $\Gamma$). They have
also been widely used to study the response of the reflection spectrum and the
\fe\ line to changes in metal abundance (Matt, Fabian \& Reynolds
1997; Ballantyne, Fabian \& Ross 2002), and inclination angle (Ghisellini,
Haardt \& Matt 1994).

\section{Non-uniform density models}
\label{sect:hydro}
Although a constant density accretion disc does have some theoretical
justification (a standard thin radiation-pressure dominated accretion
disc has a roughly constant vertical density profile; Shakura \&
Sunyaev 1973), the material on the irradiated surface is not in
pressure equilibrium if the density is held fixed. Ross \etal\ (1999)
computed the reflection spectrum from a disc where the density
followed a Gaussian drop-off. The resulting spectrum exhibited much
weaker reflection features than the equivalent constant density model.

The next step in non-uniform density models was to enforce the
condition of hydrostatic equilibrium on the illuminated gas. Initial
investigations on the disc structure of illuminated hydrostatic discs
were done by R\'{o}\.{z}a\'{n}ska \& Czerny (1996) and
R\'{o}\.{z}a\'{n}ska (1999). The full radiative transfer and
photoionization problem was then performed by Nayakshin, Kazanas \&
Kallman (2000) and subsequently investigated in a series of papers
(Nayakshin \& Kallman 2001; Nayakshin \& Kazanas 2002). These
calculations found that in many cases the illuminated gas structure took up a
two-phase appearance with a hot ($\sim$10$^7$~K), ionized zone on the
surface and a colder ($\sim 10^5$~K), recombined zone deeper in the
atmosphere. The two zones would be in pressure balance with each other
and the transition between them could be very sharp due to the thermal
ionization instability (e.g., Krolik, McKee \& Tarter 1981). In this
case, the reflection spectrum would appear to be a diluted form of a
neutral reflector as all of the emission lines would arise from the
cold recombined zone and then are scattered as they pass through the outer
hot ionized skin. Therefore a neutral \fe\ line at 6.4~\kev\ would be
expected even when the disc is highly illuminated. 

\begin{figure*}
\begin{minipage}{180mm}
\centerline{
        \psfig{figure=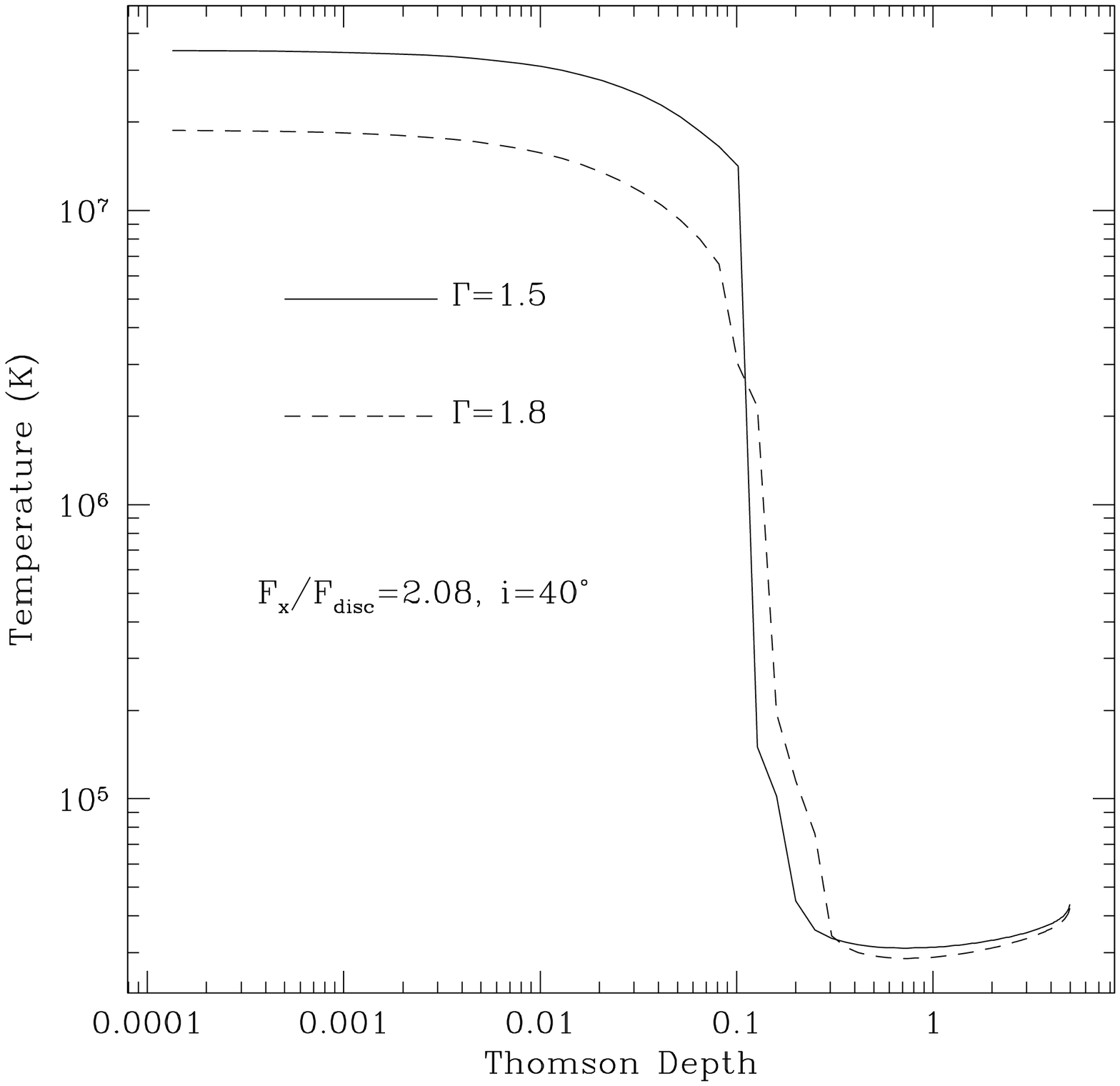,width=0.48\textwidth}
        \psfig{figure=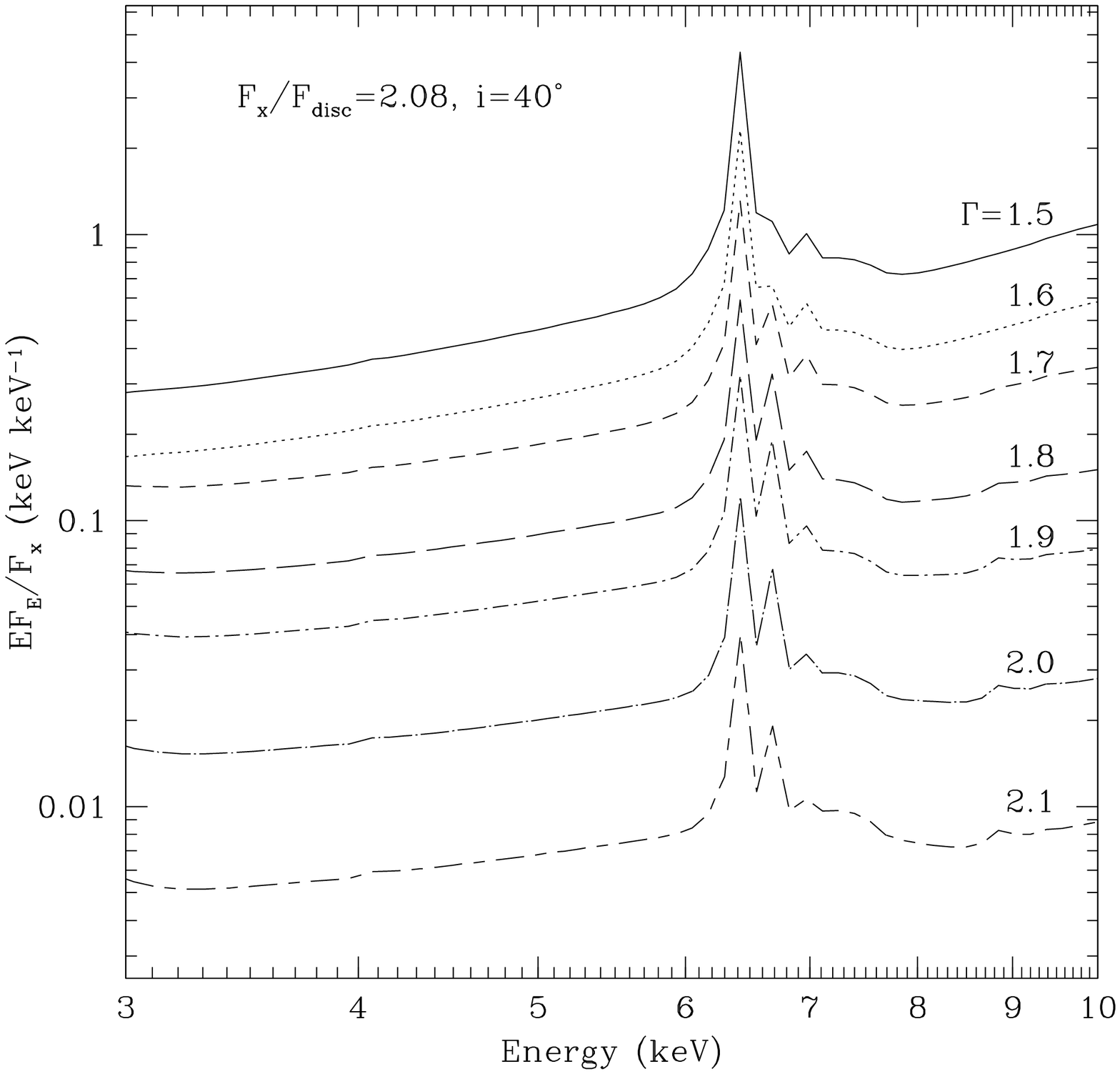,width=0.48\textwidth}
}
\caption[]{The left panel shows the temperature structure of the
        illuminated atmosphere taken from two models presented by
        Ballantyne \& Ross (2002). The sharp temperature transition
        occurs for the $\Gamma=1.5$ model, but not for the
        $\Gamma=1.8$ model which has a lower surface temperature. The
        corresponding \fe\ lines are shown in the right-hand
        panel. The $\Gamma=1.5$ model exhibits a strong neutral \fe\
        line, while the $\Gamma=1.8$ spectrum also shows a line from
        He-like Fe at 6.7~\kev. In this panel, the spectra are
        vertically offset for clarity. All the models assumed a black
        hole mass of 10$^8$~M$_{\odot}$, an accretion rate of 0.1\% of
        Eddington, and that the reflection occurred at 7 Schwarszchild
        radii from the black hole. The illuminating flux was 2.08
        times the disc flux and was incident at an angle of 40 deg
        from the normal.}
\label{fig:hydro}
\end{minipage}
\end{figure*}

The sharpness of the temperature drop between the two zones has a
great impact on the features in the reflection spectrum. Ballantyne,
Ross \& Fabian (2001) and Ballantyne \& Ross (2002) present a series of
hydrostatic calculations which show reflection spectra with strong
ionized \fe\ lines. The nearly
discontinuous drop in temperature, indicative of the thermal ionization instability,
seems to only occur when the Compton temperature of the illuminated
gas is a few keV (see Figure~\ref{fig:hydro}).
However, these considerations are subject to uncertainties in the
illuminating radiation field, particularly the high-energy cutoff of
the incident power-law (Ballantyne \& Ross 2002), and also in the true
structure of the surface of accretion discs. The question of how
important is the thermal ionization instability in determining the
reflection spectra is still one that requires more research.

While it is possible to fit X-ray data with hydrostatic models
(Nayakshin's models are available in XSPEC), they are subject
to a large number of assumptions on the accretion disc structure, and
it is this author's opinion that the constant density models are more
useful in fitting AGN data. One of the problems is that in order to
calculate a hydrostatic model, the height of the accretion disc at the
irradiated point must be known. This requires assuming a black hole
mass, accretion rate, a radius, and a full disc model before the
calculation can proceed. Therefore, these models may be more useful in
fitting data from Galactic black hole candidates where there are often
independent estimates of at least some of these parameters.

\section{Comparison with data}
\label{sect:data}
Recently, the sophisticated reflection models discussed above have
been applied to the data. This section discusses two very different
uses of the ionized constant density models. 

\begin{figure*}
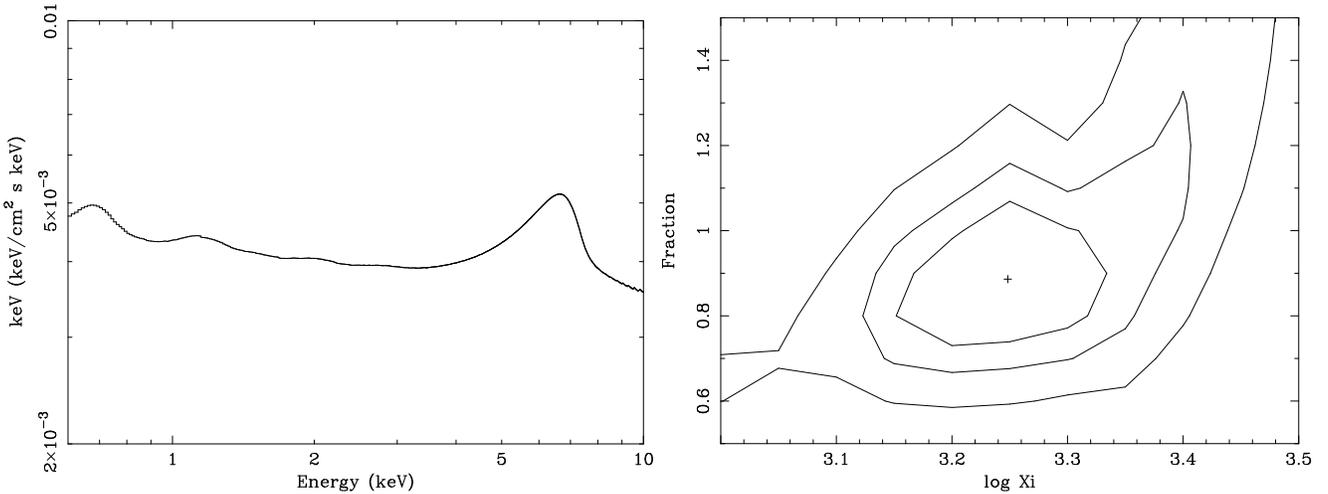

\begin{minipage}{180mm}
\centerline{
        \psfig{figure=Ballantyne_fig3a.ps,width=0.48\textwidth,angle=-90}
        \psfig{figure=Ballantyne_fig3b.ps,width=0.48\textwidth,angle=-90}
}
\caption[]{The left hand panel shows the relativistically blurred
        ionized disc model fitted to the \textit{ASCA} spectrum of the
        NLS1 Mrk~335. The right-hand panel shows confidence contours
        in the $\log \xi$-reflection fraction plane. This NLS1
        requires an ionized reflector with strong emission from a
        He-like \fe\ line at 6.7~\kev. Taken from Ballantyne, Iwasawa
\& Fabian (2001).}
\label{fig:mrk335}
\end{minipage}
\end{figure*}

Ballantyne, Iwasawa \& Fabian (2001) fitted \textit{ASCA} data of five
Narrow-Line Seyfert~1 (NLS1) galaxies with the models of Ross \&
Fabian (1993). NLS1s are thought to be systems in which a smaller than
'normal' black hole (say, 10$^6$~M$_{\odot}$) is accreting at a high
fraction of its Eddington rate (Pounds, Done \& Osborne 1995; Boller,
Brandt \& Fink 1996). In such a situation it is expected that the disc
would be hot and ionized. Observations of many NLS1s have provided some
evidence for this (Comastri \etal\ 1998, 2001; Turner, George \&
Nandra 1998, Vaughan \etal\ 1999, Turner \etal\ 2001a,b), but this
hypothesis needed to be tested with self-consistent models. Ballantyne
\etal\ (2001) found that four out of the five NLS1 considered were well fit by
the ionized reflection models (Fig.~\ref{fig:mrk335}), a stronger
constraint than measuring the energy of the \fe\ line because the
continuum is fit simultaneously with the line.

\begin{figure}
\centerline{\psfig{figure=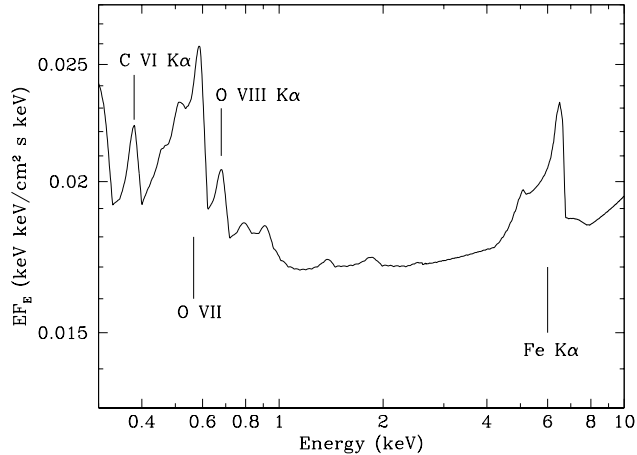,width=0.48\textwidth,angle=-90}}
\caption[]{This plot shows the best fit constant density reflection model
  determined from fitting the 1994 \textit{ASCA} data of \mcg\ above
  3~\kev. The prominent emission lines are indicated. The model
  predicts that the O~\textsc{viii} emission line will have an EW of only
5.4~eV. From Ballantyne \& Fabian (2001).}
\label{fig:mcg6-model}
\end{figure}

Recent \textit{XMM-Newton} observations of \mcg\ and Mrk~766 by
Branduardi-Raymont \etal\ (2001) suggested that there were strong relativistic
emission lines from C, N and O in the soft X-ray part of the
spectrum. Ionized reflection models can provide predictions on the
strength of such lines as they are predicted along with \fe\ and the
reflection continuum. Ballantyne \& Fabian (2001) fit the
\textit{ASCA} data of \mcg\ around the \fe\ line with both constant
density and hydrostatic reflection models. The models then predicted
the strength and shape of the carbon and oxygen lines (see
Fig.~\ref{fig:mcg6-model}).
Both types of models predicted that the EWs of the soft X-ray lines
should be on the order of tens of eV, and that O~\textsc{vii} and Fe~L
emission should both be common in the soft X-ray band. These results
raise questions about the relativistic diskline interpretation of
Branduardi-Raymont \etal\ (2001).

The current most common reflection models in use in XSPEC are
\textsc{pexrav} for neutral reflection and \textsc{pexriv} for ionized
reflection (Magdziarz \& Zdziarski 1995). Both are constant density
models, and both calculate the reflection continuum, including
edges. This means that emission lines, such as \fe\ must be added in
separately to the model which is not an ideal situation because, as
Fig.~\ref{fig:rfy} shows, the properties of the line are closely
linked with the reflection continuum. For ionized reflection,
\textsc{pexriv} suffers from inaccuracies at high ionization
parameters because it neglects Comptonization (see Fig.~7 in Ross
\etal\ 1999). A better alternative for comparing reflection models to
data is to use the constant density ones computed using the code of
Ross \& Fabian (1993). Two grids of these models (one with solar Fe
abundance, the other at twice solar Fe) are now available for public
use as XSPEC table models\footnote{From
\texttt{http://legacy.gsfc.nasa.gov/docs/xanadu/xspec/models/iondisc.html}}.
The grids cover a wide range of parameter space ($1.0 \leq \log \xi
\leq 6.0$, $1.5 \leq \Gamma \leq 3.0$, $0.0 \leq R \leq 2$, where $R$
is the reflection fraction: total=$R\times$reflected+incident) and of
course include many emission lines computed self-consistently with the
reflection continuum.

\begin{acknowledgements}
DRB acknowledges financial support from the Commonwealth Scholarship and
Fellowship Plan and the Natural Sciences and Engineering Research
Council of Canada. ACF and RRR acknowledge support from the Royal
Society and the College of the Holy Cross, respectively.
\end{acknowledgements}

\end{document}